\documentclass[11pt,a4paper]{article}
\usepackage[hyperref]{acl2020}
\usepackage{times}
\usepackage{latexsym}

\usepackage{textcomp}

\usepackage{graphicx}
\usepackage{tabularx}
\usepackage{soul}

\usepackage{url}

\usepackage{epstopdf}
\usepackage[latin1]{inputenc}

\usepackage{hyperref}
\usepackage{xstring}
\usepackage{enumitem} 

\usepackage{cleveref}

\usepackage{xcolor}
\usepackage{epstopdf}
\newcommand{\bl}{\color{black}}

\usepackage{microtype}

\aclfinalcopy 

\title{Gender Gap in Natural Language Processing Research:\\
Disparities in Authorship and Citations
}

\author{Saif M. Mohammad\\
National Research Council Canada\\
Ottawa, Canada\\ 
\texttt{saif.mohammad@nrc-cnrc.gc.ca}\\}

\date{}

\begin{document}
\maketitle
\begin{abstract}
Disparities in authorship and citations across gender can have substantial
adverse consequences not just on the disadvantaged genders, but also on the field of study as a whole. 
{\bl Measuring gender gaps is a crucial step towards addressing them.}
In this work, we examine female first author percentages and the citations to their papers in Natural Language Processing {\bl (1965 to 2019)}.
{\bl We determine aggregate-level statistics using an existing manually curated 
author--gender list as well as first names strongly associated with a gender.}
We find that only about 29\% of first authors are female and only about 25\% of last authors are female. Notably, this percentage has not improved since the mid 2000s. We also show that, on average, female first authors are cited less than male first authors, even when controlling for experience and area of research. 
{\bl Finally, we discuss the ethical considerations involved in automatic
demographic analysis.} 
\end{abstract}


\section{Introduction}
\setitemize[0]{leftmargin=*}
\setenumerate[0]{leftmargin=*}

Gender gaps are quantitative measures of the disparities 
in social, political, intellectual, cultural, or economic success due to one's gender. 
Gender gaps can also refer to disparities in access to resources (such as healthcare and education), 
which in turn lead to disparities in success.
We need to pay attention to gender gaps not only because they are inherently unfair but also
because better gender balance leads to higher productivity, better health and well-being, greater economic benefits,
better decision  making, as well as political and economic stability \cite{skjelsboek2001gender,woetzel2015power,hakura2016inequality,mehta2017gender,gallego2018integrated}. 

{\bl 
Historically, gender has often been considered binary (male and female), immutable (cannot change), and physiological (mapped to biological sex). 
However, those views have been discredited \cite{hyde2019future,richards2017non,darwin2017doing,lindsey2015sociology,kessler1978gender}. 
Gender is complex, and does not necessarily fall into binary male or female categories (e.g. nonbinary people), and also does not necessarily correspond to one's assigned gender at birth.

Society has often viewed different gender groups differently, imposing unequal social and power structures \cite{lindsey2015sociology}.}
The World Economic Forum's 2018 Global Gender Gap Report (which examined data from more than 144 countries) highlighted the gender gap between men and women in Artificial Intelligence  as particularly
alarming \cite{GGG18}. 
It indicated that only 22\% of the professionals in AI are women and that this low representation in a transformative field 
requires urgent action---otherwise, the AI gap has the potential to widen other gender gaps.
Other studies have identified substantial gender gaps in science \cite{haakanson2005impact, lariviere2013bibliometrics, king2017men, andersen2018google}.

{\bl \citet{perez2019invisible} discusses, through numerous examples, how there is a considerable lack
of disaggregated data for women and how that is directly leading to negative outcomes in all spheres of their lives, including health,
income, safety, and the degree to which they succeed in their endeavors. This holds true even more for transgender people. 
Our work obtains disaggregated data for female Natural Language Processing (NLP) researchers and determines the degree of gender gap between female and male NLP researchers.}
(NLP is an interdisciplinary field that includes scholarly work on language and computation
with influences from Artificial Intelligence, Computer Science, Linguistics, Psychology, and Social Sciences to name a few.)
{\bl We hope future work will explore other gender gaps (e.g.,
between trans and cis people). Measuring gender gaps is a crucial step towards addressing them.}


We examine tens of thousands of articles in the ACL Anthology (AA) 
(a digital repository of public domain NLP articles)
for disparities in female authorship.\footnote{https://www.aclweb.org/anthology/} 
We also conduct experiments to determine whether female first authors are cited more or less than male first authors,
using citation counts extracted  from Google Scholar (GS).

We extracted and aligned information from the ACL Anthology and Google Scholar to create a dataset of tens of thousands of NLP papers and their citations as part of a broader project on analyzing NLP Literature.\footnote{\citet{mohammad2019nlpscholar} presents an overview of the many research directions pursued, using this data.
Notably, \citet{mohammad2020citations} explores questions such as: how well cited are papers of different types (journal articles, conference papers, demo papers, etc.)?
how well cited are papers published in different time spans?
how well cited are papers from different areas of research within NLP? etc. \citet{mohammad2020demo} presents an interactive visualization tool that allows users to search for relevant related work in the ACL Anthology.} 
We refer to this dataset as the \textit{NLP Scholar Dataset}. 
{\bl We determined aggregate-level statistics for female and male researchers 
in the NLP Scholar dataset
using an existing manually curated 
 author--gender list as well as first names that are strongly associated with a gender.}

{\bl Note that attempts to automatically infer gender of individuals
can lead to harm \cite{hamidi2018gender}.
Our work does not aim to infer gender of individual authors. We 
use name--gender association information
to determine aggregate-level statistics for male and female researchers.
Further, one may not know most researchers they cite, other than from reading their work. 
Thus perceived gender (from the name) can lead to unconscious effects,
e.g., \citet{dion2018gendered} show that 
all male and mixed author teams cite fewer papers by female authors
than all female teams.
Further, seeing only a small number of female authors cited can demoralize 
young researchers entering the field.} 

We do not explore the reasons behind gender gaps. 
However, we will note that the reasons are often complex, intersectional, and difficult to disentangle. 
We hope that this work will increase awareness of gender gaps and inspire concrete steps to improve inclusiveness and fairness in research. 

It should also be noted that even though this paper focuses on {\bl female--male disparities}, 
there are many aspects to demographic diversity including: representation from {\bl transgender people}; representation from various nationalities and race; representation by people who speak a diverse set of languages; diversity by income, age, physical abilities, etc. All of these factors impact the breadth of technologies we create, how useful they are, and whether they reach those that need it most.

Resources for the NLP Scholar project can be accessed through the project homepage.\footnote{http://saifmohammad.com/WebPages/nlpscholar.html}

\section{Related Work}

{\bl \citet{pilcher2017names} shows that names function not only to identify individuals, but also
to manage gender throughout one's life. There is a strong cultural norm in various parts of the world to assign a first name to newborns as per their category of sex \cite{pilcher2017names,  barry2014unisex, lieberson2000instability, alford1987naming}. 
\citet{pilcher2017names} argues that, throughout their life, the first name plays a role in repeatedly categorizing a person as being male or female. People may change their name and appearance to manage their gender \cite{connell2010doing,pilcher2017names}.
People may choose a name that is not associated with male or female categorizations \cite{connell2010doing}.

The strong normative tendency to use names to signal gender
has led to a large body of work on automatically determining gender by one's first name,
not just for scientometric analysis discussed below, but also for language studies, social sciences, public health, and commerce.
However, this can also lead to \textit{misgendering}, which can cause significant pain and harm \cite{hamidi2018gender}. (Misgendering is when a person---or in this case, a machine---associates someone with a gender with which they do not identify.)
Further, work that does not explicitly consider gender to be inclusive of trans people can reinforce stereotypes such as the dichotomy of gender.

We expect gender disparities to be different depending on the groups being compared: female--male,
trans--cis, and so on.
Our work does not aim to infer gender of individual authors. 
We obtain disaggregated statistics for women, specifically, to study the disparities
between female and male NLP researchers.
We discuss ethical considerations further in Section \ref{sec:ethics}.   
See also \citet{mihaljevic2019reflections} for a discussion on ethical considerations 
in using author name to estimate gender statistics in 
the Gender Gap in Science Project---a large ongoing project tracking gender gaps in 
Mathematical and Natural Sciences.}\footnote{https://gender-gap-in-science.org}

Most studies on gender and authorship 
have found substantial gender disparities in favor of male researchers. They include work on
$\sim$1700 articles from journals of library and information science \cite{haakanson2005impact},  
on $\sim$12 million articles from the Web of Science (for Sociology, Political Science, Economics, Cardiology and Chemistry) \cite{ghiasi2016gender,andersen2018google},        
on $\sim$2 million mathematics articles \cite{mihaljevic2016effect},
on $\sim$1.6 million articles from PubMed life science and biomedical research  \cite{mishra2018self},
on $\sim$1.5 million articles from fifty disciplines published in JSTOR \cite{king2017men}, 
and on $\sim$0.5 million publications from US research universities \cite{duch2012possible}. 
There also exists some work that shows that in fields such as linguistics \cite{LSA17} and psychology \cite{willyard2011men},
female and male participation is either close to parity or tilted in favor of women. 

In NLP research,
\newcite{schluter2018glass} showed 
that there are barriers in the paths of women researchers,
delaying their attainment of  mentorship status (as estimated through last author position in papers).
\newcite{anderson2012towards} examine papers from 1980 to 2008 to track 
  the ebb and flow of topics within NLP,
 and the influence of researchers from outside NLP on NLP.
\newcite{vogel2012he} examined about 13,000 papers from 1980 to 2008
to determine basic authorship statistics by women and men. Gender statistics were determined
by a combination of automatic and manual means. The automatic method relied
on lists of baby names from various languages.
They found that female authorship has been steadily increasing from 1980 to 2008.
Our work examines a much larger set of NLP papers (1965--2019), re-examines some of the questions
raised in \newcite{vogel2012he}, and explores several new questions, especially on
first author gender and
disparities in citation. 

\section{Data}
We extracted and aligned information from the ACL Anthology (AA) and Google Scholar (GS) to create a dataset of tens of thousands of 
NLP papers and their citations.
We aligned the information across AA and GS using the paper title, year of publication, and first author last name.
Details about the dataset, as well as an analysis of the volume of  research in NLP over the years,
are available in \citet{mohammad2020data}.
We summarize key information below. 

\subsection{ACL Anthology Data}
\label{AAsec}

The ACL Anthology 
is available through its website and a \texttt{github} repository.\footnote{https://www.aclweb.org/anthology/\\https://github.com/acl-org/acl-anthology}
We extracted paper title, names of authors, year of publication, and venue of publication from the repository.\footnote{Multiple authors can have the same name and the same authors may use multiple variants of their names in papers. 
The AA volunteer team handles such ambiguities using both semi-automatic and manual approaches (fixing some instances on a case-by-case basis). 
Additionally, AA keeps a file that includes canonical forms of author names.}


As of June 2019, AA had $\sim$50K entries; however, this includes some entries that are not truly research publications (for example, forewords, prefaces, programs, schedules, indexes, invited talks, appendices, session information, newsletters, lists of proceedings, etc.). 
After discarding them, we are left with 44,894 papers.\footnote{We used simple keyword searches for terms such as
 {\it foreword, invited talk, program, appendix} and {\it session} in the title to pull out
 entries that were likely to not be research publications. These were then manually examined to verify
 that they did not contain any false positives.}\\[-8pt]


\noindent {\bf \bl Inferring Aggregate Gender Statistics:} 
The ACL Anthology
does not record author demographic information. 
{\bl To infer aggregate statistics for male and female authors,
we create two bins of authors: A-Mname (authors that have self-reported as males or with names commonly associated with males)
and A-Fnames (authors that have self-reported as females or with names commonly associated with females).\footnote{Note that a person may have a name commonly
associated with one gender but belong to a different gender.} 
}
We made use of three resources to {\bl populate A-Mname and A-Fname}:\\[-16pt]
\begin{enumerate}
\item A manually curated list of 11,932 AA authors and their genders provided by 
\citet{vogel2012he} (VJ-AA list) (3,359 female and 8,573 male).\footnote{https://nlp.stanford.edu/projects/gender.shtml}  \\[-18pt]
\item A list of 55,924 first names that are strongly associated with females and 30,982 first names that are strongly associated with males, that we generated from the US Social Security Administration's (USSA) published database of names and genders of newborns.\footnote{https://www.ssa.gov/oact/babynames/limits.html}  \\[-18pt]
\item A list of 26,847 first names that are strongly associated with females and 23,614 first names that are strongly associated with males, that we generated from
a list of 9,300,182 PUBMED authors and their genders \cite{torvik2009author,smith2013search}.\footnote{https://experts.illinois.edu/en/datasets/genni-ethnea-for-the-author-ity-2009-dataset-2}\\[-16pt]
\end{enumerate}



\noindent  We acknowledge that despite a large expatriate population, the US census information is not
representative of the names from around the world. Further, Chinese origin names tend not to be as strongly associated with gender as names from other parts of the world. 
However, it should be noted that \citet{vogel2012he} made special effort to include information from a large number of 
Asian AA authors in their list.
The PUBMED list is also noted for having a substantial coverage of Asian names \cite{torvik2009author}.

\begin{table}
{\small
\begin{small}
    	\begin{center}
	    \begin{tabular}{lrrr}\hline
	       firstname--gender association list     &P      &R      &F\\\hline
            from USSA data   &98.4   &69.8   &81.7\\
            from PUBMED data &98.3   &81.4   &89.1\\
	\hline
		\end{tabular}
	\end{center}
	\end{small}
	\caption{\label{tab:namesF} Precision (P), Recall (R), and F-score (F) of how well the first name and gender association matches information in the VJ-AA list.}
	}
\end{table}

We determined first name--gender association, by calculating the percentages of first names corresponding to male and female genders as per each of the PUBMED and USSA fullname--gender lists.
We consider a first name to be strongly associated with a gender if the percentage is $\geq 95$\%.\footnote{A choice of other percentages such as 90\% or 99\% would also have been reasonable.}
Table \ref{tab:namesF} shows {\bl how well the first name and gender association matches with the VJ-AA list}.


Given the high precision (over 98\%) of the USSA and PUBMED lists of gender-associated first names, we use them (in addition to the VJ-AA list)
{\bl to populate the M-names and F-names bins}. 
Eventually, {\bl the A-Mname and A-Fname bins} together had 28,682 (76\%) of the 37,733 AA authors.
{\bl Similarly, we created bins for Papers whose First Author is from A-Mname (P-FA-Mname),
Papers whose First Author is from A-Fname (P-FA-Fname),
Papers whose Last Author is from A-Mname (P-LA-Mname), and
Papers whose Last Author is from A-Fname (P-LA-Fname)
to estimate aggregate-level statistics for papers with male and female first and last authors.
} {\bl P-FA-Mname and P-FA-Fname together} have 37,297 (83\%) AA papers (we will refer to this subset as AA*),
{\bl P-LA-Mname and P-LA-Fname together} have 39,368 (88\%) AA papers (we will refer to this subset  as AA**). 

\noindent {\bf NLP Academic Age as a Proxy for Experience in NLP:} 
First author percentage may vary due to 
experience, area of research within NLP, venue of publication, etc.
To gauge experience, we use the number of years one has been publishing in AA; we will refer to to this as the \textit{NLP Academic Age}.   
So if this is the first year one has published in AA, then their NLP academic age is 1. If one published their first AA paper in 2001 and their latest AA paper in 2018, then their academic age is 18.
Note that NLP academic age is not always an accurate reflection of one's research experience.
Also, one can publish NLP papers outside of AA. 

\subsection{Google Scholar Data}
\label{sec:GS}

Google Scholar 
allows researchers to create and edit public author profiles called \textit{Google Scholar Profiles}. Authors can include their papers (along with their citation information) on this page.

We extracted citation information from Google Scholar profiles of authors who published at least three papers in the ACL Anthology.\footnote{This is  allowed by GS's robots exclusion standard.} 
This yielded citation information for 1.1 million papers in total. We will refer to this dataset as the \textit{NLP Subset of the Google Scholar Dataset}, or \textit{GScholar-NLP} for short. Note that GScholar-NLP includes citation counts not just for NLP papers, but also for non-NLP papers published by authors who have at least three papers in AA.

GScholar-NLP includes 32,985 
of the 44,894 papers in AA (about 75\%). We will refer to this subset of the ACL Anthology papers as AA$'$. The citation analyses presented in this paper are on AA$'$.

\section{Gender Gap in Authorship}

{\bl We use the 
datasets described in \S\ref{AAsec} (especially AA* and AA**)} to answer a series of questions
on female authorship. 
{\bl Since we do not have full self-reported information for all authors, these should be treated as estimates.}

First author is a privileged position in the author list that is usually reserved for the researcher that has done the most work and writing.\footnote{A small number of papers have more than one first author. This work did not track that.}
In NLP, first authors are also often students. 
Thus we are especially interested in gender gaps that effect them.
The last author position is often reserved for the most senior or mentoring researcher. 
We explore last author disparities only briefly (in Q1).\\[-8pt]



\noindent \textit{Q1. What percentage of the authors in AA are female? What percentage of the AA papers have female first authors (FFA)?
What percentage of the AA papers have female last authors (FLA)?
How have these percentages changed since 1965?}\\[-8pt]

\begin{figure*}[t!]
 \begin{center}
 	\includegraphics[width=2\columnwidth]{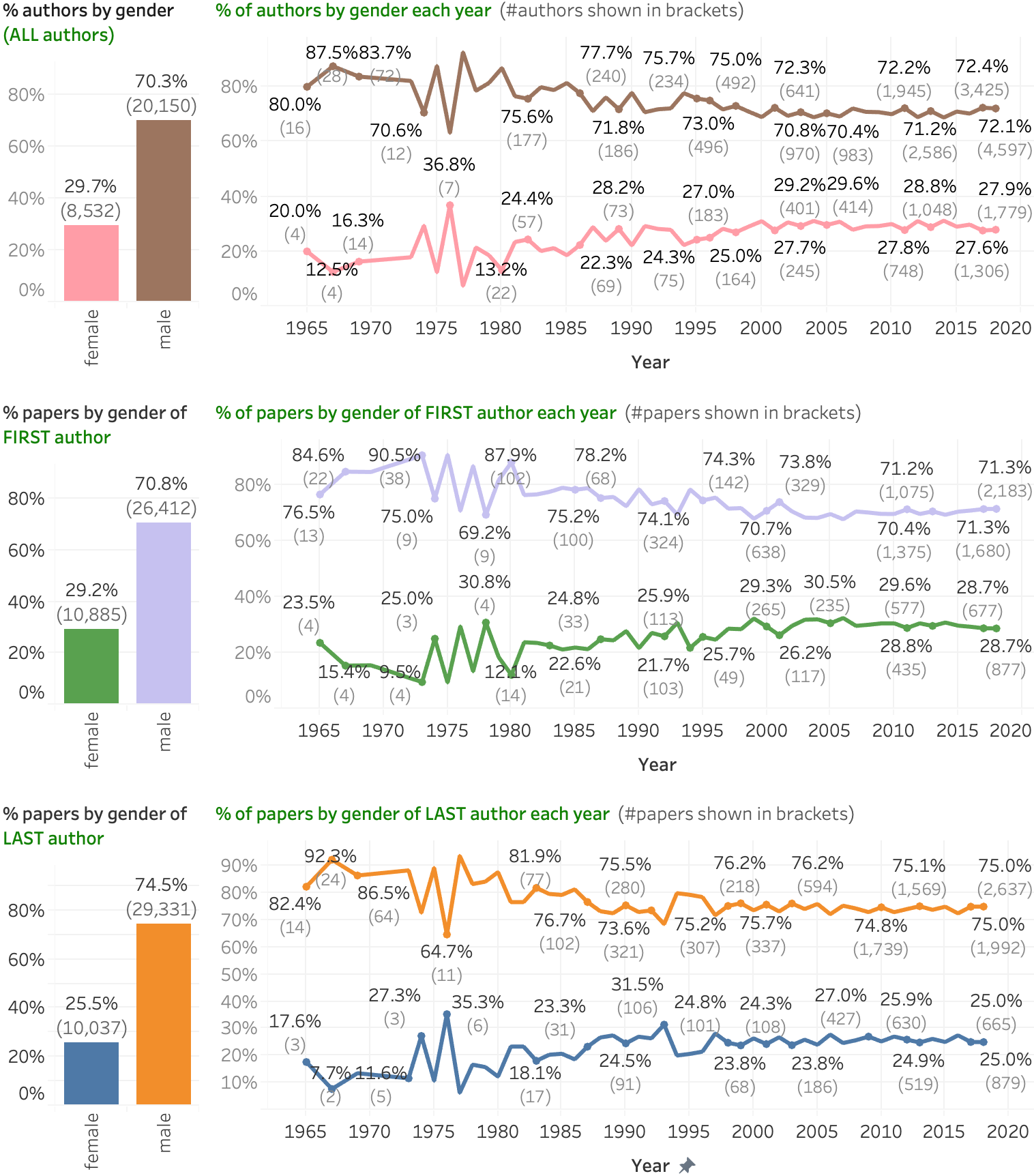}
\vspace*{-1mm}
 	\caption{Female authorship percentages in AA over the years: overall, as first author, and as last author.}
 	\label{fig:FFAP}
 \end{center}
\vspace*{-1mm}
 \end{figure*}

\noindent A. 
Overall, {\bl we estimate that} about 29.7\% of the 28,682 authors  
are female; 
about 29.2\% of the first authors in 37,297 AA* papers 
are female; 
and about 25.5\% of the last authors in 39,368 AA** papers 
are female.
Figure \ref{fig:FFAP} shows how these percentages have changed over the years.\\[-10pt] 

\noindent \textit{Discussion:} Across the years, the percentage of female authors overall is close to the percentage of papers with female first authors. (These percentages are around 28\% and 29\%, respectively, in 2018.) However, the percentage of female last authors is markedly lower (hovering at about 25\% in 2018).
These numbers indicate that, as a community, we are far from obtaining male--female parity.
A further striking (and concerning) observation is that the female author percentages have not improved since the year 2006.



To put these numbers in context, the percentage of female scientists worldwide (considering all areas of research) has been estimated to be around 30\%. The reported percentages for many computer science sub-fields are much lower.\footnote{https://unesdoc.unesco.org/ark:/48223/pf0000235155}
The percentages are much higher for certain other fields such as psychology \cite{willyard2011men} and linguistics \cite{LSA17}.\\[-8pt]

   \begin{figure*}[t!]
 \begin{center}
 	\includegraphics[width=2\columnwidth]{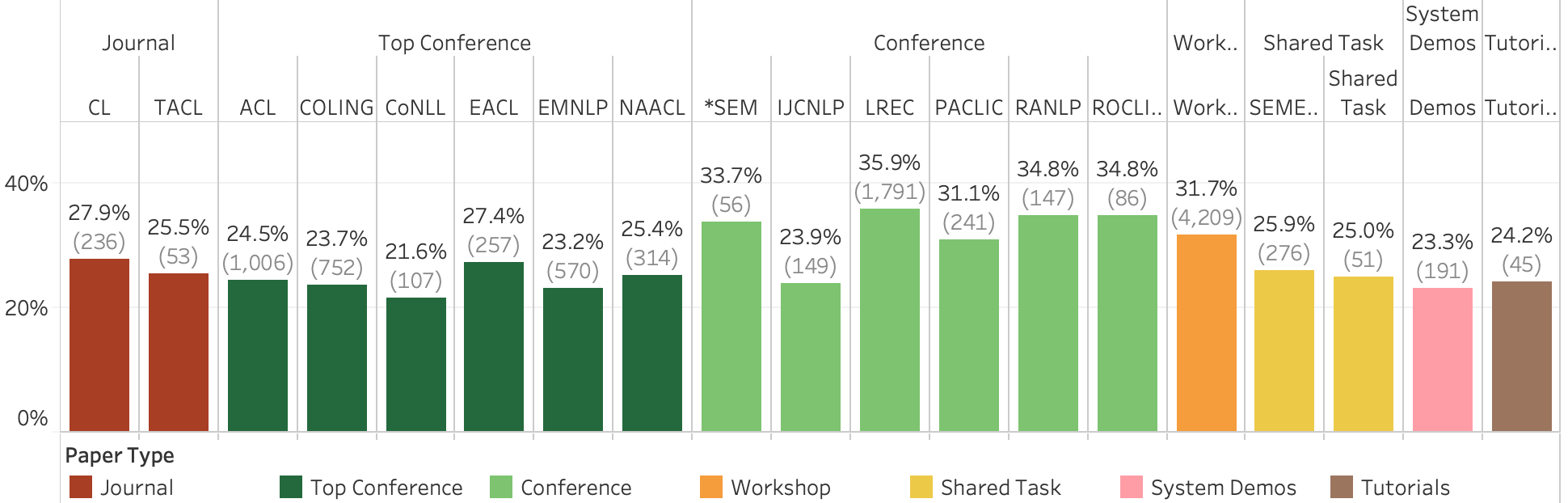}
 	\caption{FFA percentage by venue and paper type. The number of FFA papers is shown in parenthesis.}
 	\label{fig:FFAP-venues}
 \end{center}
 \vspace*{-3mm}
 \end{figure*}

 \noindent \textit{Q2. How does FFA vary by paper type and venue?}\\[-8pt]

\noindent A. Figure \ref{fig:FFAP-venues} shows FFA percentages by paper type and venue. \\[-10pt]

\noindent \textit{Discussion:} Observe that FFA percentages are lowest for CoNLL, EMNLP, IJCNLP, and
system demonstration papers (21\% to 24\%). FFA percentages for journals, other top-tier conferences, 
SemEval, shared task papers, and tutorials are the next lowest (24\% to 28\%). 
The percentages are markedly higher for LREC, *Sem, and RANLP (33\% to 36\%), as well as for workshops (31.7\%).\\[-8pt]

   \begin{figure}[t!]
 \begin{center}
 	\includegraphics[width=\columnwidth]{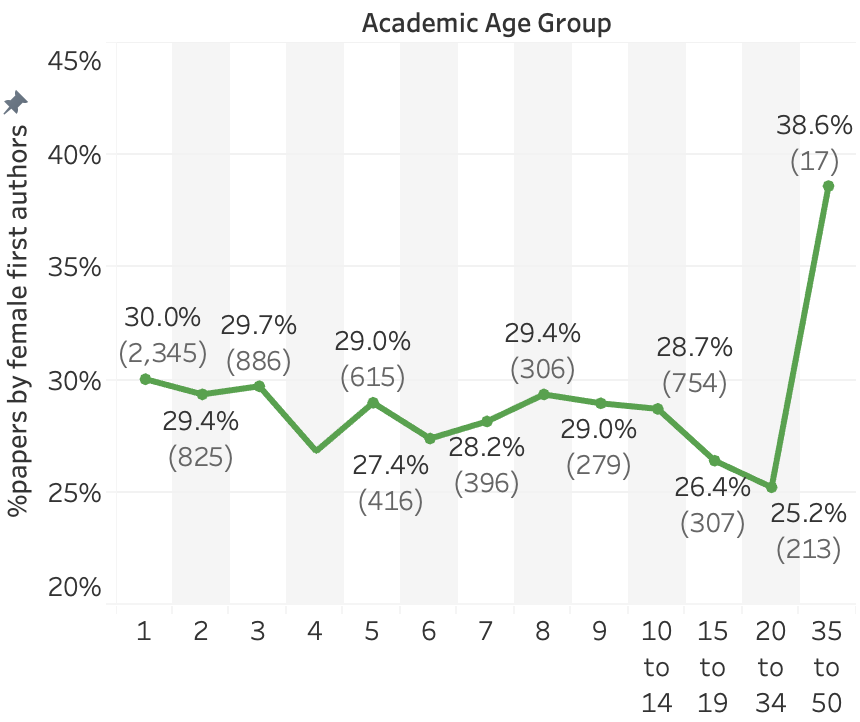}
 	\caption{FFA percentage by academic age. The number of FFA papers is shown in parenthesis. }
 	\label{fig:FFAP-AAge}
 \end{center}
 \vspace*{-3mm}
 \end{figure}

\noindent \textit{Q3. How does female first author percentage change with NLP academic age?}\\[-8pt]

\noindent A.  In order to determine these numbers, every paper in AA* was placed in a bin corresponding to NLP academic age: if the paper's first author had an academic age of 1 in the year when the paper was published, then the paper is placed in bin 1; if the paper's first author had an academic age of 2 in the year when the paper was published, then the paper is placed in bin 2; and so on.  The bins for later years contained fewer papers. This is expected as senior authors in NLP often work with students, and students are encouraged to be first authors.
Thus, we combine some of the bins in later years: one bin for academic ages between 10 and 14; one for 15 to 19; one for 20 to 34; and one for 35 to 50.
Once the papers are assigned to the bins, we calculate the percentage of papers in each bin that have a female first author.
Figure \ref{fig:FFAP-AAge} shows the results.\\[-10pt]

\noindent \textit{Discussion:} Observe that, with the exception of the 35 to 50 academic age bin, FFA\% is highest (30\%) 
at age 1 (first year of publication).
There is a period of decline in FFA\% until year 6 (27.4\%)---this difference is statistically significant (t-test, p $<$ 0.01). 
This might be a potential indicator that graduate school has a progressively greater negative impact on the productivity of women than of men. (Academic age 1 to 6 often correspond to the period when the first author is in graduate school or in a temporary post-doctoral position.)
After year 6, we see a recovery back to 29.4\% by year 8, followed by a period of decline once again. \\[-8pt]


\noindent \textit{Q4. How does female first author percentage vary by area of research (within NLP)? Which areas have higher-than-average FFA\%? Which areas have lower-than-average FFA\%? How does FFA\% correlate with popularity of an area---that is, does FFA\% tend to be higher- or lower-than-average in areas where lots of authors are publishing?}\\[-10pt]

    \begin{figure}[t!]
 \begin{center}
 	\includegraphics[width=0.925\columnwidth]{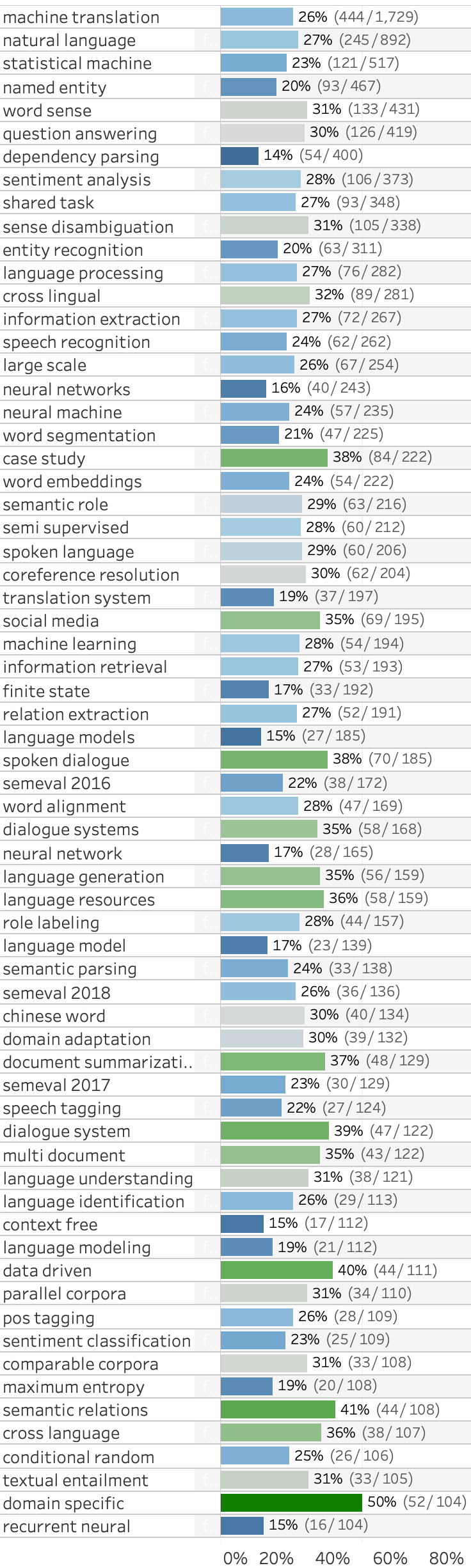}
 	 \vspace*{-2mm}
 	\caption{Top 66 bigrams in AA* titles and FFA\%
 	(30\%: light grey; $<$30\%: blue; $>$30\%: green).}
 	\label{fig:FFA-bigrams75}
 \end{center}
 \vspace*{-10mm}
 \end{figure}
 
 
\noindent A. We use word bigrams in the titles of papers to sample papers from various areas.\footnote{Other approaches such as clustering are also reasonable; however, results with those might not be easily reproducible.} 
The title has a privileged position in a paper. 
Primarily, it conveys what the paper is about.
For example, a paper with {\it machine translation} in the title is likely about machine translation.
Figure \ref{fig:FFA-bigrams75} shows the list of top 66 bigrams that occur in the titles of more than 100 AA* papers (in decreasing order of the bigram frequency). 
For each bigram, the figure also shows the percentage of papers with a female first author. 
In order to determine whether there is a correlation between the number of papers corresponding to a bigram and FFA\%, we calculated the Spearman's rank correlation between the rank of a bigram by number of papers and the rank of a bigram by FFA\%. This correlation was found to be only 0.16. This correlation is not statistically significant at p $<$ 0.01 (two-sided p-value = 0.2). 
Experiments with lower thresholds 
(174 bigrams occurring in 50 or more papers and 1408 bigrams occurring in 10 or more papers) also resulted in very low and non-significant correlation numbers (0.11 and 0.03, respectively).\\[-12pt]

\noindent \textit{Discussion:} Observe that FFA\% varies substantially depending on the bigram. It is particularly low for title bigrams such as \textit{dependency parsing, language models, finite state, context free,} and \textit{neural models}; and markedly higher than average for \textit{domain specific, semantic relations, dialogue system, spoken dialogue, document summarization,} and \textit{language resources}.
However, the rank correlation experiments show that there is \textit{no} correlation between the popularity of an area (number of papers that have a bigram in the title) and the percentage of female first authors.  

To obtain further insights, we also repeat some of the experiments described above for unigrams in paper titles. 
We found that FFA rates are relatively high in non-English European language research such as papers on Russian, Portuguese, French, and Italian.
FFA rates are also relatively high for 
work on prosody, readability, discourse, dialogue, paraphrasing, and individual parts of speech such as adjectives and verbs.
FFA rates are particularly low for papers on theoretical aspects of statistical modelling, and 
for terms such as \textit{WMT, parsing, markov, recurrent}, and \textit{discriminative}.



\section{Gender Gap in Citations}

Research articles can have impact in a number of ways---pushing the state of the art, answering crucial questions, finding practical solutions that directly help people, etc.
However, individual measures of research impact are limited in scope---they measure only some kinds of contributions.
The most commonly used metrics of research impact are derived from citations including: number of citations, average citations, h-index, 
and impact factor \cite{bornmann2009state}.
Despite their limitations, citation metrics have substantial impact on a researcher's scientific career; often through a combination of funding, the ability to attract talented students and collaborators, job prospects, and other opportunities in the wider research community.
Thus, disparities in  citations (citation gaps) across demographic attributes such as gender, race, and location 
have direct real-world adverse implications. This often also results in the demoralization of researchers and marginalization of their work---thus negatively impacting the whole field. 

Therefore, we examine gender disparities in citations in NLP.  We use 
a subset of the 32,985 AA$'$ papers (\S\ref{sec:GS}) that were published from 1965 to 2016 for the analysis (to allow for at least 2.5 years for the papers to collect citations). There are 26,949 such papers.\\[-10pt]


 \begin{figure}[t!]
 \begin{center}
\includegraphics[width=0.78\columnwidth]{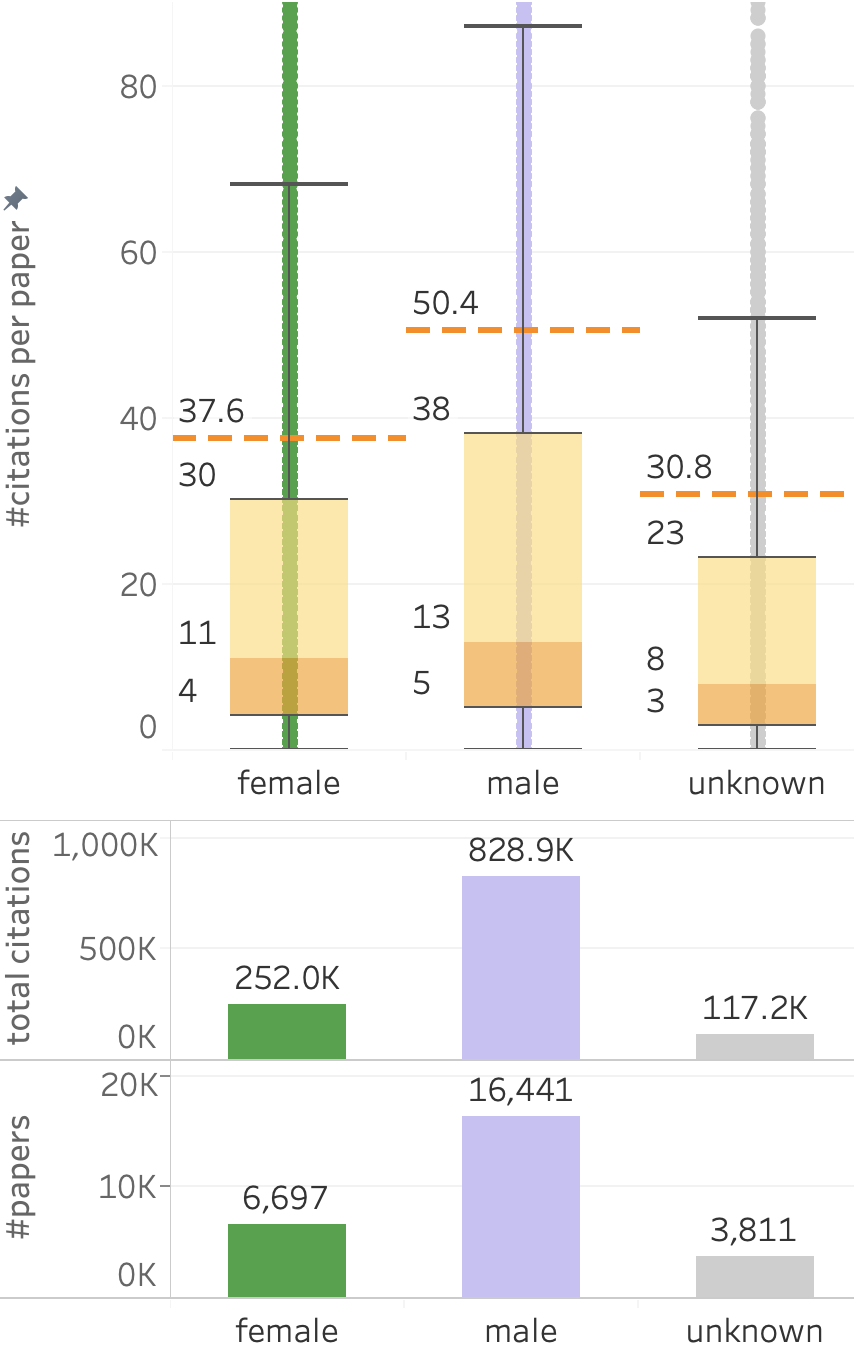}
 	\caption{\#papers, total citations, box plot of citations per paper: for female, male, gender-unknown first authors. The orange dashed lines mark averages.}
 	\label{fig:citns-gender}
 \end{center}
\vspace*{-5mm}
 \end{figure}
 
  \begin{figure*}[t!]
 \begin{center}
 \includegraphics[width=1.88\columnwidth]{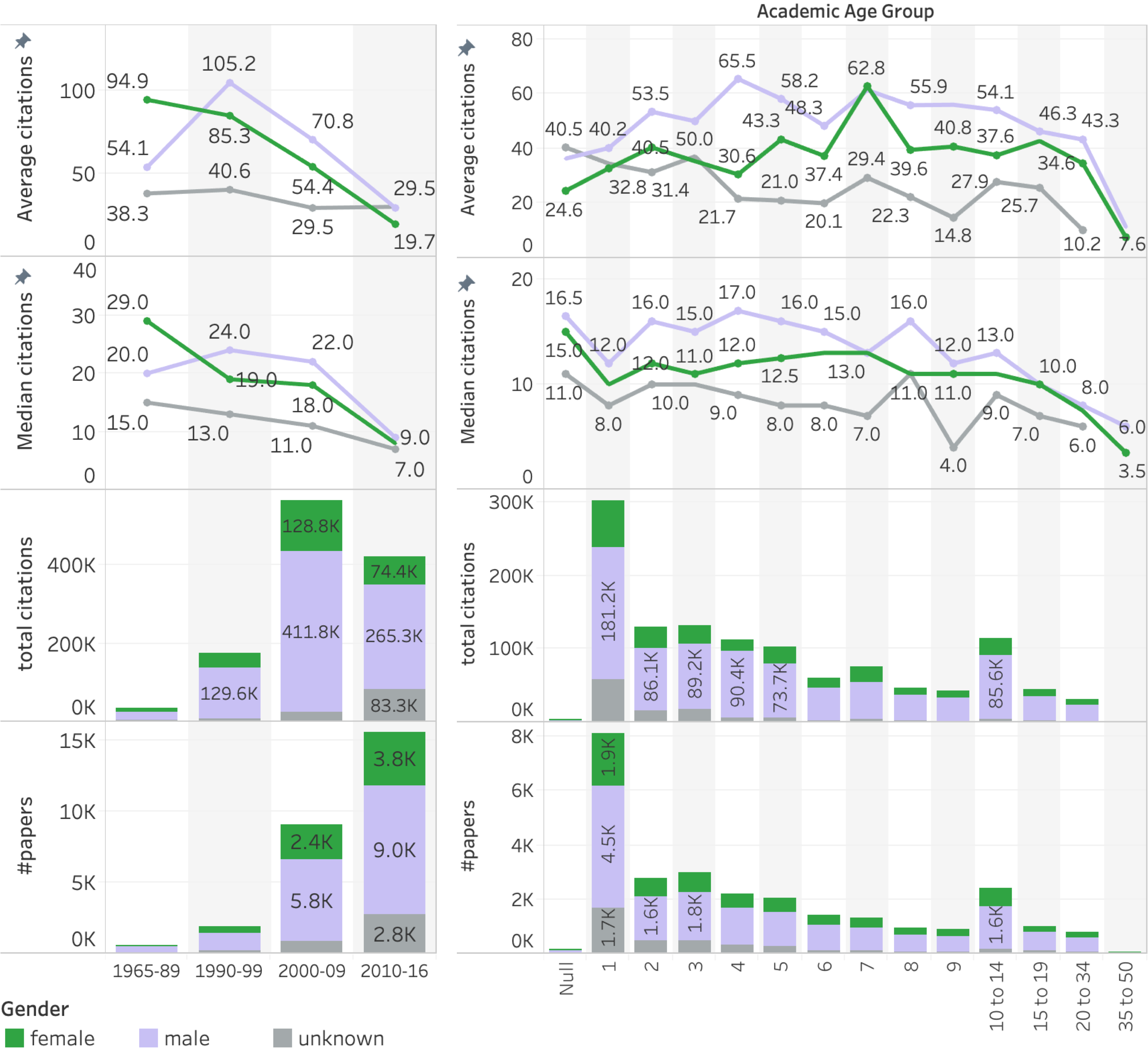}
 \vspace*{2mm}
 	\caption{Citation gap across genders for papers: published in different time spans (left); by academic age (right).}
 	\label{fig:citn-gender-time-band}
 \end{center}
 \vspace*{-2mm}
 \end{figure*}
 
\noindent \textit{Q5. How well cited are women and men?} \\[-10pt] 

\noindent A. For all three classes (females, males, and gender unknown), Figure \ref{fig:citns-gender} shows: a bar graph of number of papers, a bar graph of total citations received, and box and whisker plots for citations received by individuals.
The whiskers are at a distance of 1.5 times the inter-quartile length. 
Number of citations pertaining to key points such as 25th percentile, median, and 75th percentile are indicated on the left of the corresponding horizontal bars. \\[-12pt]

\noindent \textit{Discussion:} On average, female first author papers have received markedly fewer citations than male first author papers (37.6 compared to 50.4). The difference in median is smaller (11 compared to 13).
The difference in the distributions of males and females is statistically significant (Kolmogorov--Smirnov test, p $<$ 0.01 ).\footnote{Kolmogorov--Smirnov (KS) test is a non-parametric test that can be applied to compare any two distributions without making assumptions about the nature of the distributions.} 
The large difference in averages and smaller difference in medians suggests that there are markedly more very heavily cited male first-author papers than female first-author papers.\\[-10pt]

\noindent The differences in citations, or \textit{citation gap}, across genders may itself vary: (1) by period of time; (2) due to confounding factors such as academic age and areas of research. We explore these next.\\[-10pt]

\noindent \textit{Q6. How has the citation gap across genders changed over the years?}\\[-10pt]

\noindent A. Figure \ref{fig:citn-gender-time-band} (left side) shows the citation statistics across four time periods.\\[-12pt]

\noindent \textit{Discussion:} Observe that female first authors have always been a minority in the history of ACL; however, on average, their papers from the early years (1965 to 1989) received a markedly higher number of citations than those of male first authors from the same period. We can see from the graph that this changed in the 1990s when male first-author papers obtained markedly more citations on average. The citation gap reduced considerably in the 2000s, and the 2010--2016 period saw a further reduction. 
It remains to be seen whether the citation gap for these 2010--2016 papers widens in the coming years. 

It is also interesting to note that the gender-unknown category has almost bridged the gap with the male category in terms of average citations. 
Further, the proportion of the gender-unknown authors has steadily increased over the years---arguably, an indication of better representation of authors from around the world in recent years.\footnote{Our method is expected to have a lower coverage of names from outside North America and Europe because USSA and PUBMED databases historically have had fewer names from outside North America and Europe.}\\[-10pt] 

\noindent \textit{Q7. How have citations varied by gender and academic age? Is the  citation gap a side effect of a greater proportion of new-to-NLP female first authors than new-to-NLP male first authors?}\\[-10pt]

\noindent A. Figure \ref{fig:citn-gender-time-band} (right side) shows citation statistics broken down by gender and academic age.\\[-12pt]

\noindent \textit{Discussion:} The graphs show that female first authors consistently receive fewer citations than male  first authors across the spans of their academic age. (The gap is highest at academic age 4 and lowest at academic age 7.) Thus,
the citation gap is likely due to factors beyond differences in average academic age between men and women.\\[-10pt]




\noindent \textit{Q8. How prevalent is the citation gap across areas of research within NLP? Is the gap simply because more women work in areas that receive low numbers of citations (regardless of gender)?}\\[-10pt]

\noindent A. On average, male first authors are cited more than female first authors in 54 of the 66 areas (82\% of the areas) discussed earlier in Q4 and Figure \ref{fig:FFA-bigrams75}. 
Female first authors are cited more in the sets of papers whose titles have: 
\textit{word sense, sentiment analysis, 
information extraction,
neural networks, neural network,
semeval 2016, language model, document summarization, multi document, spoken dialogue, dialogue systems,
} and {\it speech tagging}. 

If women chose to work in areas that happen to attract less citations by virtue of the area, then we would not expect to see citation gaps in so many areas.
Recall also that we already showed that FFA\% is not correlated with rank of popularity of an area (Q4). 
Thus it is 
unlikely that the choice of area of research is behind the gender gap.

{\bl 
\section{Limitations and Ethical Considerations}
\label{sec:ethics}

\noindent \textit{Q9. What are the limitations and ethical considerations involved with this work?}\\[-10pt]

\noindent A. Data is often a representation of people \cite{zook2017ten}. This is certainly the case here and we acknowledge that the use of such data has the potential to harm individuals. 
Further, while the methods used are not new, their use merits reflection. 

Analysis focused on women and men leaves out non-binary people.\footnote{Note that as per widely cited definitions, nonbinary people are considered transgender, but most transgender people are not non-binary. Also, trans people often use a name that is more associated with their gender identity.} 
Additionally, not disaggregating cis and trans people means that the statistics are largely reflective of the more populous cis class.
We hope future work will obtain disaggregated information for various genders. However, careful attention must be paid as
some gender classes might include too few NLP researchers to ensure anonymity even with aggregate-level analysis.

The use of female- and male-gender associated names to infer population level statistics for women and men 
  excludes people that do not have such names and people from some cultures where names are not as strongly associated with gender.

Names are not immutable
(they can be changed to indicate or not indicate gender)
and 
people can choose to keep their birth name or change it (providing autonomy). 
However, changing names can be quite difficult.
Also, names do not capture gender fluidity or contextual gender. 
   
   A more inclusive way of obtaining gender information is through self-reported surveys. However, 
   challenges persist in terms of how to design effective and inclusive questionnaires \cite{bauer2017transgender,geniuss2014best}.
   Further, even with self-report textboxes that give the respondent the primacy and autonomy to express gender, downstream research often ignores such data or combines information in ways beyond the control of the respondent.\footnote{https://reallifemag.com/counting-the-countless/} Also, as is the case here, it is not easy to obtain self-reported historical information. 

Social category detection can potentially lead to harm, for example, depriving people of opportunities simply because of their race or gender. However, one can also see the benefits of NLP techniques and social category detection in public health (e.g., developing targeted initiatives to improve health outcomes of vulnerable populations), as well as in psychology and social science (e.g., to better understand the unique challenges of belonging to a social category). 


A larger list of ethical considerations associated with the NLP Scholar project is available through the project webpage.\footnote{http://saifmohammad.com/WebPages/nlpscholar.html}
\citet{mihaljevic2019reflections} also discusses the ethical considerations in using
author names to infer gender statistics in the Gender Gap in Science Project.\footnote{https://gender-gap-in-science.org}}

\section{Conclusions}
We analyzed 
the ACL Anthology to show that only $\sim$30\% have female authors, $\sim$29\% have female first authors, and $\sim$25\% have female last authors. Strikingly, even though some gains were made in the early years of NLP, 
overall FFA\% has not improved since the mid 2000s.
Even though there are some areas where FFA\% is close to parity with male first authorship, most areas have a substantial gap in the numbers for male and female authorship. We found no correlation between popularity of research area and FFA\%. 
We also showed how FFA\% varied by paper type, venue, academic age, and area of research. We used citation counts extracted from Google Scholar to show that,
on average, 
male first authors are cited markedly more than female first authors, even when controlling for experience and area of work. 
Thus, in NLP,  gender gaps exist both in authorship and citations.

This paper did not explore the reasons behind the gender gaps.
However, 
the inequities that impact the number of women pursuing scientific research  \cite{roos2008together,foschi2004blocking,buchmann2009gender} 
and biases that impact citation patterns unfairly \cite{brouns2007making,feller2004measurement,gupta2005triple} are well-documented. 
These factors play a substantial role in creating the gender gap,
as opposed to differences in innate ability or  
differences in quality of work produced by these two genders. If anything, 
past research has shown that self-selection in the face of inequities and adversity leads to more competitive, capable, and confident cohorts \cite{nekby2008gender,hardies2013gender}. 

\section*{Acknowledgments}

\vspace*{-2mm}
Many thanks to Rebecca Knowles, Ellen Riloff, Tara Small, Isar Nejadgholi, Dan Jurafsky, Rada Mihalcea, Isabelle Augenstein, Eric Joanis, Michael Strube, Shubhanshu Mishra, and Cyril Goutte for the tremendously helpful discussions.
{\bl Many thanks to Cassidy Rae Henry, Luca Soldaini, Su Lin Blodgett, Graeme Hirst, Brendan T. O'Connor, Luc\'ia Santamar\'ia, Lyle Ungar, Emma Manning, and Peter Turney for
discussions on the ethical considerations involved with this work.
}\\[-20pt]

\bibliography{ACL2020-Gender_Gap_in_NLP}

\begin{thebibliography}{51}
\expandafter\ifx\csname natexlab\endcsname\relax\def\natexlab#1{#1}\fi

\bibitem[{Alford(1987)}]{alford1987naming}
Richard Alford. 1987.
\newblock \emph{Naming and identity: A cross-cultural study of personal naming
  practices}.
\newblock Hraf Press.

\bibitem[{Andersen and Nielsen(2018)}]{andersen2018google}
Jens~Peter Andersen and Mathias~Wullum Nielsen. 2018.
\newblock {Google Scholar and Web of Science}: Examining gender differences in
  citation coverage across five scientific disciplines.
\newblock \emph{Journal of Informetrics}, 12(3):950--959.

\bibitem[{Anderson et~al.(2012)Anderson, McFarland, and
  Jurafsky}]{anderson2012towards}
Ashton Anderson, Dan McFarland, and Dan Jurafsky. 2012.
\newblock Towards a computational history of the {ACL}: 1980--2008.
\newblock In \emph{Proceedings of the Workshop on Rediscovering 50 Years of
  Discoveries}, pages 13--21.

\bibitem[{Barry~III and Harper(2014)}]{barry2014unisex}
Herbert Barry~III and Aylene~S Harper. 2014.
\newblock Unisex names for babies born in pennsylvania 1990--2010.
\newblock \emph{Names}, 62(1):13--22.

\bibitem[{Bauer et~al.(2017)Bauer, Braimoh, Scheim, and
  Dharma}]{bauer2017transgender}
Greta~R Bauer, Jessica Braimoh, Ayden~I Scheim, and Christoffer Dharma. 2017.
\newblock Transgender-inclusive measures of sex/gender for population surveys:
  Mixed-methods evaluation and recommendations.
\newblock \emph{PloS one}, 12(5):e0178043.

\bibitem[{Bornmann and Daniel(2009)}]{bornmann2009state}
Lutz Bornmann and Hans-Dieter Daniel. 2009.
\newblock The state of h index research.
\newblock \emph{EMBO reports}, 10(1):2--6.

\bibitem[{Brouns(2007)}]{brouns2007making}
Margo Brouns. 2007.
\newblock The making of excellence: {G}ender bias in academia.
\newblock In \emph{Exzellenz in Wissenschaft und Forschung - neue Wege in der
  Gleichstellungspolitik}, pages 23--42. Wissenshaftsrat.

\bibitem[{Buchmann(2009)}]{buchmann2009gender}
Claudia Buchmann. 2009.
\newblock Gender inequalities in the transition to college.
\newblock \emph{Teachers College Record}, 111(10):2320--2345.

\bibitem[{Connell(2010)}]{connell2010doing}
Catherine Connell. 2010.
\newblock Doing, undoing, or redoing gender? {L}earning from the workplace
  experiences of transpeople.
\newblock \emph{Gender \& Society}, 24(1):31--55.

\bibitem[{Darwin(2017)}]{darwin2017doing}
Helana Darwin. 2017.
\newblock Doing gender beyond the binary: A virtual ethnography.
\newblock \emph{Symbolic Interaction}, 40(3):317--334.

\bibitem[{Dion et~al.(2018)Dion, Sumner, and Mitchell}]{dion2018gendered}
Michelle~L Dion, Jane Sumner, and Sara~McLaughlin Mitchell. 2018.
\newblock Gendered citation patterns across political science and social
  science methodology fields.
\newblock \emph{Political Analysis}, 26(3):312--327.

\bibitem[{Duch et~al.(2012)Duch, Zeng, Sales-Pardo, Radicchi, Otis, Woodruff,
  and Amaral}]{duch2012possible}
Jordi Duch, Xiao Han~T Zeng, Marta Sales-Pardo, Filippo Radicchi, Shayna Otis,
  Teresa~K Woodruff, and Lu{\'\i}s A~Nunes Amaral. 2012.
\newblock The possible role of resource requirements and academic career-choice
  risk on gender differences in publication rate and impact.
\newblock \emph{PloS one}, 7(12):e51332.

\bibitem[{Feller(2004)}]{feller2004measurement}
Irwin Feller. 2004.
\newblock Measurement of scientific performance and gender bias.
\newblock In \emph{Gender and Excellence in the Making}, pages 35--39.
  Luxembourg: Office for Official Publications of the European Communities.

\bibitem[{Foschi(2004)}]{foschi2004blocking}
Marta Foschi. 2004.
\newblock Blocking the use of gender-based double standards for competence.
\newblock In \emph{Gender and Excellence in the Making}, pages 51--55.
  Luxembourg: Office for Official Publications of the European Communities.

\bibitem[{Gallego and Guti{\'e}rrez(2018)}]{gallego2018integrated}
Juan~Miguel Gallego and Luis~H Guti{\'e}rrez. 2018.
\newblock An integrated analysis of the impact of gender diversity on
  innovation and productivity in manufacturing firms.
\newblock Technical report, Inter-American Development Bank.

\bibitem[{Ghiasi et~al.(2016)Ghiasi, Larivi{\`e}re, and
  Sugimoto}]{ghiasi2016gender}
Gita Ghiasi, Vincent Larivi{\`e}re, and Cassidy Sugimoto. 2016.
\newblock Gender differences in synchronous and diachronous self-citations.
\newblock In \emph{21st International Conference on Science and Technology
  Indicators-STI 2016. Book of Proceedings}.

\bibitem[{Group(2014)}]{geniuss2014best}
GenIUSS Group. 2014.
\newblock \emph{Best practices for asking questions to identify transgender and
  other gender minority respondents on population-based surveys}.
\newblock eScholarship, University of California.

\bibitem[{Gupta et~al.(2005)Gupta, Kemelgor, Fuchs, and
  Etzkowitz}]{gupta2005triple}
Namrata Gupta, Carol Kemelgor, Stefan Fuchs, and Henry Etzkowitz. 2005.
\newblock Triple burden on women in science: A cross-cultural analysis.
\newblock \emph{Current science}, pages 1382--1386.

\bibitem[{H{\aa}kanson(2005)}]{haakanson2005impact}
Malin H{\aa}kanson. 2005.
\newblock The impact of gender on citations: An analysis of college \& research
  libraries, journal of academic librarianship, and library quarterly.
\newblock \emph{College \& Research Libraries}, 66(4):312--323.

\bibitem[{Hakura et~al.(2016)Hakura, Hussain, Newiak, Thakoor, and
  Yang}]{hakura2016inequality}
Dalia~S Hakura, Mumtaz Hussain, Monique Newiak, Vimal Thakoor, and Fan Yang.
  2016.
\newblock \emph{Inequality, gender gaps and economic growth: Comparative
  evidence for sub-Saharan Africa}.
\newblock International Monetary Fund.

\bibitem[{Hamidi et~al.(2018)Hamidi, Scheuerman, and
  Branham}]{hamidi2018gender}
Foad Hamidi, Morgan~Klaus Scheuerman, and Stacy~M Branham. 2018.
\newblock Gender recognition or gender reductionism? {T}he social implications
  of embedded gender recognition systems.
\newblock In \emph{Proceedings of the 2018 {CHI} conference on human factors in
  computing systems}, pages 1--13.

\bibitem[{Hardies et~al.(2013)Hardies, Breesch, and
  Branson}]{hardies2013gender}
Kris Hardies, Diane Breesch, and Jo{\"e}l Branson. 2013.
\newblock Gender differences in overconfidence and risk taking: Do
  self-selection and socialization matter?
\newblock \emph{Economics Letters}, 118(3):442--444.

\bibitem[{Hyde et~al.(2019)Hyde, Bigler, Joel, Tate, and van
  Anders}]{hyde2019future}
Janet~Shibley Hyde, Rebecca~S Bigler, Daphna Joel, Charlotte~Chucky Tate, and
  Sari~M van Anders. 2019.
\newblock The future of sex and gender in psychology: Five challenges to the
  gender binary.
\newblock \emph{American Psychologist}, 74(2):171.

\bibitem[{Kessler and McKenna(1978)}]{kessler1978gender}
Suzanne~J Kessler and Wendy McKenna. 1978.
\newblock \emph{Gender: An ethnomethodological approach}.
\newblock IL: The University of Chicago Press.

\bibitem[{King et~al.(2017)King, Bergstrom, Correll, Jacquet, and
  West}]{king2017men}
Molly~M King, Carl~T Bergstrom, Shelley~J Correll, Jennifer Jacquet, and
  Jevin~D West. 2017.
\newblock Men set their own cites high: Gender and self-citation across fields
  and over time.
\newblock \emph{Socius}, 3:2378023117738903.

\bibitem[{Larivi{\`e}re et~al.(2013)Larivi{\`e}re, Ni, Gingras, Cronin, and
  Sugimoto}]{lariviere2013bibliometrics}
Vincent Larivi{\`e}re, Chaoqun Ni, Yves Gingras, Blaise Cronin, and Cassidy~R
  Sugimoto. 2013.
\newblock Bibliometrics: Global gender disparities in science.
\newblock \emph{Nature News}, 504(7479):211.

\bibitem[{Lieberson et~al.(2000)Lieberson, Dumais, and
  Baumann}]{lieberson2000instability}
Stanley Lieberson, Susan Dumais, and Shyon Baumann. 2000.
\newblock The instability of androgynous names: The symbolic maintenance of
  gender boundaries.
\newblock \emph{American Journal of Sociology}, 105(5):1249--1287.

\bibitem[{Lindsey(2015)}]{lindsey2015sociology}
Linda~L Lindsey. 2015.
\newblock The sociology of gender theoretical perspectives and feminist
  frameworks.
\newblock In \emph{Gender roles}, pages 23--48. Routledge.

\bibitem[{LSA(2017)}]{LSA17}
The Linguistic Society of~America LSA. 2017.
\newblock The state of linguistics in higher education annual report 2017.
\newblock Technical report, The Linguistic Society of America.

\bibitem[{Mehta et~al.(2017)Mehta, Burns, Machado, Fox-Robichaud, Cook, Calfee,
  Ware, Burnham, Kissoon, Marshall et~al.}]{mehta2017gender}
Sangeeta Mehta, Karen~EA Burns, Flavia~R Machado, Alison~E Fox-Robichaud,
  Deborah~J Cook, Carolyn~S Calfee, Lorraine~B Ware, Ellen~L Burnham, Niranjan
  Kissoon, John~C Marshall, et~al. 2017.
\newblock Gender parity in critical care medicine.
\newblock \emph{American journal of respiratory and critical care medicine},
  196(4):425--429.

\bibitem[{Mihaljevi{\'c} et~al.(2019)Mihaljevi{\'c}, Tullney, Santamar{\'\i}a,
  and Steinfeldt}]{mihaljevic2019reflections}
Helena Mihaljevi{\'c}, Marco Tullney, Luc{\'\i}a Santamar{\'\i}a, and Christian
  Steinfeldt. 2019.
\newblock Reflections on gender analyses of bibliographic corpora.
\newblock \emph{Frontiers in Big Data}, 2:29.

\bibitem[{Mihaljevi{\'c}-Brandt et~al.(2016)Mihaljevi{\'c}-Brandt,
  Santamar{\'\i}a, and Tullney}]{mihaljevic2016effect}
Helena Mihaljevi{\'c}-Brandt, Luc{\'\i}a Santamar{\'\i}a, and Marco Tullney.
  2016.
\newblock The effect of gender in the publication patterns in mathematics.
\newblock \emph{PLoS One}, 11(10):e0165367.

\bibitem[{Mishra et~al.(2018)Mishra, Fegley, Diesner, and
  Torvik}]{mishra2018self}
Shubhanshu Mishra, Brent~D Fegley, Jana Diesner, and Vetle~I Torvik. 2018.
\newblock Self-citation is the hallmark of productive authors, of any gender.
\newblock \emph{PloS one}, 13(9):e0195773.

\bibitem[{Mohammad(2019)}]{mohammad2019nlpscholar}
Saif~M. Mohammad. 2019.
\newblock The state of {NLP} literature: A diachronic analysis of the {ACL
  Anthology}.
\newblock \emph{arXiv preprint arXiv:1911.03562}.

\bibitem[{Mohammad(2020{\natexlab{a}})}]{mohammad2020citations}
Saif~M. Mohammad. 2020{\natexlab{a}}.
\newblock Examining citations of natural language processing literature.
\newblock In \emph{Proceedings of the 2020 Annual Conference of the Association
  for Computational Linguistics}, Seattle, USA.

\bibitem[{Mohammad(2020{\natexlab{b}})}]{mohammad2020data}
Saif~M. Mohammad. 2020{\natexlab{b}}.
\newblock {NLP Scholar}: A dataset for examining the state of {NLP} research.
\newblock In \emph{Proceedings of the 12th Language Resources and Evaluation
  Conference (LREC-2020)}, Marseille, France.

\bibitem[{Mohammad(2020{\natexlab{c}})}]{mohammad2020demo}
Saif~M. Mohammad. 2020{\natexlab{c}}.
\newblock {NLP Scholar}: An interactive visual explorer for {Natural Language
  Processing} literature.
\newblock In \emph{Proceedings of the 2020 Annual Conference of the Association
  for Computational Linguistics}, Seattle, USA.

\bibitem[{Nekby et~al.(2008)Nekby, Thoursie, and Vahtrik}]{nekby2008gender}
Lena Nekby, Peter Thoursie, and Lars Vahtrik. 2008.
\newblock Gender and self-selection into a competitive environment: Are women
  more overconfident than men?
\newblock \emph{Economics Letters}, 100(3):405--407.

\bibitem[{Perez(2019)}]{perez2019invisible}
Caroline~Criado Perez. 2019.
\newblock \emph{Invisible women: Exposing data bias in a world designed for
  men}.
\newblock Random House.

\bibitem[{Pilcher(2017)}]{pilcher2017names}
Jane Pilcher. 2017.
\newblock Names and ``doing gender{"}: How forenames and surnames contribute to
  gender identities, difference, and inequalities.
\newblock \emph{Sex roles}, 77(11-12):812--822.

\bibitem[{Richards et~al.(2017)Richards, Bouman, and Barker}]{richards2017non}
Cristina Richards, Walter~Pierre Bouman, and M~Barker. 2017.
\newblock Non-binary genders.
\newblock \emph{London: Pal grave Macmillan}.

\bibitem[{Roos(2008)}]{roos2008together}
Patricia~A Roos. 2008.
\newblock Together but unequal: Combating gender inequity in the academy.
\newblock \emph{Journal of Workplace Rights}, 13(2):185--199.

\bibitem[{Schluter(2018)}]{schluter2018glass}
Natalie Schluter. 2018.
\newblock \href {https://doi.org/10.18653/v1/D18-1301} {The glass ceiling in
  {NLP}}.
\newblock In \emph{Proceedings of the 2018 Conference on Empirical Methods in
  Natural Language Processing}, pages 2793--2798, Brussels, Belgium.

\bibitem[{Skjelsboek and Smith(2001)}]{skjelsboek2001gender}
Inger Skjelsboek and Dan Smith. 2001.
\newblock \emph{Gender, peace and conflict}.
\newblock Sage.

\bibitem[{Smith et~al.(2013)Smith, Singh, and Torvik}]{smith2013search}
Brittany~N Smith, Mamta Singh, and Vetle~I Torvik. 2013.
\newblock A search engine approach to estimating temporal changes in gender
  orientation of first names.
\newblock In \emph{Proceedings of the 13th ACM/IEEE-CS joint conference on
  Digital libraries}, pages 199--208. ACM.

\bibitem[{Torvik and Smalheiser(2009)}]{torvik2009author}
Vetle~I Torvik and Neil~R Smalheiser. 2009.
\newblock Author name disambiguation in medline.
\newblock \emph{ACM Transactions on Knowledge Discovery from Data (TKDD)},
  3(3):1--29.

\bibitem[{Vogel and Jurafsky(2012)}]{vogel2012he}
Adam Vogel and Dan Jurafsky. 2012.
\newblock He said, she said: Gender in the {ACL Anthology}.
\newblock In \emph{Proceedings of the Special Workshop on Rediscovering 50
  Years of Discoveries}, pages 33--41, Jeju Island, Korea.

\bibitem[{WEF(2018)}]{GGG18}
World Economic~Forum WEF. 2018.
\newblock The global gender gap report 2018.
\newblock Technical report, World Economic Forum, Geneva, Switzerland.

\bibitem[{Willyard(2011)}]{willyard2011men}
Cassandra Willyard. 2011.
\newblock Men: A growing minority.
\newblock \emph{GradPSYCH Magazine}, 9(1):40.

\bibitem[{Woetzel et~al.(2015)}]{woetzel2015power}
Jonathan Woetzel et~al. 2015.
\newblock The power of parity: How advancing women's equality can add \$12
  trillion to global growth.
\newblock Technical report, McKinsey Global Institute.

\bibitem[{Zook et~al.(2017)Zook, Barocas, Boyd, Crawford, Keller, Gangadharan,
  Goodman, Hollander, Koenig, Metcalf, Narayanan, Nelson, and
  Pasquale}]{zook2017ten}
Matthew Zook, Solon Barocas, Danah Boyd, Kate Crawford, Emily Keller,
  Seeta~Pena Gangadharan, Alyssa Goodman, Rachelle Hollander, Barbara~A.
  Koenig, Jacob Metcalf, Arvind Narayanan, Alondra Nelson, and Frank Pasquale.
  2017.
\newblock \href {https://doi.org/10.1371/journal.pcbi.1005399} {Ten simple
  rules for responsible big data research}.
\newblock \emph{PLOS Computational Biology}, 13(3):1--10.

\end{thebibliography}
\bibliographystyle{acl_natbib}

\end{document}